\documentclass[ amsmath, amssym, aps, pra, groupedaddress, superscriptaddress,showkeys]{revtex4-2}

\usepackage{graphicx}
\usepackage{dcolumn}
\usepackage{bm}
\usepackage[utf8]{inputenc}     
\usepackage[T1, T2A]{fontenc}
\usepackage{amsmath,amssymb}
\usepackage[english]{babel} 
\usepackage{makecell}

\begin{document}
\title{Superconducting properties of Nb$_{0.85}$Sc$_{0.15}$ film deposited by magnetron co-sputtering} 
\author{P.A. Berezhnoy}
\affiliation{Institute of Nanotechnology of Microelectronics, Russian Academy of Sciences (INME RAS), Moscow 119991, Russia}
\affiliation{National Research Centre 'Kurchatov Institute', Moscow, Russia}

\author{A.A. Elistratova}
\affiliation{Institute of Nanotechnology of Microelectronics, Russian Academy of Sciences (INME RAS), Moscow 119991, Russia}
\affiliation{Dukhov All-Russia Research Institute of Automatics, Moscow 101000, Russia}

\author{Z.S. Enbaev}
\affiliation{Institute of Nanotechnology of Microelectronics, Russian Academy of Sciences (INME RAS), Moscow 119991, Russia}

\author{M.A. Dryazgov}
\affiliation{Institute of Nanotechnology of Microelectronics, Russian Academy of Sciences (INME RAS), Moscow 119991, Russia}

\author{A.M. Mumlyakov}
\affiliation{Institute of Nanotechnology of Microelectronics, Russian Academy of Sciences (INME RAS), Moscow 119991, Russia}

\author{O.A. Solovyev}
\affiliation{Institute of Nanotechnology of Microelectronics, Russian Academy of Sciences (INME RAS), Moscow 119991, Russia}

\author{M.O. Geodakyan}
\affiliation{Institute of Nanotechnology of Microelectronics, Russian Academy of Sciences (INME RAS), Moscow 119991, Russia}

\author{I.V. Trofimov}
\affiliation{Institute of Nanotechnology of Microelectronics, Russian Academy of Sciences (INME RAS), Moscow 119991, Russia}

\author{V.S. Stolyarov}
\affiliation{Dukhov All-Russia Research Institute of Automatics, Moscow 101000, Russia}

\author{A.A. Korneev}
\affiliation{Institute of Nanotechnology of Microelectronics, Russian Academy of Sciences (INME RAS), Moscow 119991, Russia}
\affiliation{HSE University, Moscow, 101000, Russia}  

\author{M.A. Tarkhov}
\affiliation{Institute of Nanotechnology of Microelectronics, Russian Academy of Sciences (INME RAS), Moscow 119991, Russia}
\affiliation{National Research Centre 'Kurchatov Institute', Moscow, Russia}

\begin{abstract}
The technology has been developed for synthesizing Nb$_{1-x}$Sc$_x$ films using magnetron co-sputtering from Nb and Sc targets.
The material synthesis was accompanied by structural characterization using X-ray diffraction and X-ray reflectometry methods, which enabled the determination of thickness, phase composition and crystal structure. We also analyzed the superconducting properties. 
The critical temperature $T_c$ was measured for samples with different concentarions of Sc and Nb. The maximum value of $T_c$ equal to 6.35 K was observed for sample with a scandium content of approximately 15\%,  which was determined by Auger spectroscopy. Transport and magnetoresistive measurements were performed in microbridges with length of 50 $\mu$m, width of 2 $\mu$m, and thickness of 30 nm. The critical current density was as high as 2.5 $\text{ MA/cm}^{2}$. Magnetic measurements were performed with the field oriented perpendicular to the sample. The upper critical field $H_{c2}(0) = 3.2$ T, electron diffusion coefficient $D = 1.1$ $\text{cm}^{2}/s$, and coherence length $\xi_{GL} =  10.1$ nm. The synthesized Nb$_{1-x}$Sc$_x$ intermetallic compound shows promise for various functional cryogenic electronics devices. 
\end{abstract}

\maketitle

\section{\label{sec:level1}Introduction}

In modern cryoelectroniс devices superconducting materials based on niobium (Nb) play a key role. Pure niobium stands out among elemental superconductors with the highest critical transition temperature: for bulk samples and high-purity thin films, the critical temperatures are $T_c \sim 9.2 - 9.35$ K \cite{Superconductivity1963,Tanatar2022Anisotropic,Joshi2022Quasiparticle}. Currently, pure niobium is widely used in sensitive magnetosensors, radiofrequency resonators, and as a metallization layer for superconducting qubit circuits \cite{Tanatar2022Anisotropic,Ciovati2015,Joshi2022Quasiparticle}. Another popular material is niobium nitride (NbN) with a transition temperature close to 15 K. NbN thin films exhibit high critical current densities up to 5–10 MA/$\text{cm}^2$ and upper critical fields $H_{c2}(0) > 20$ T \cite{Iilin_2008,Ilin2014,ShibalovM_2022}. This is one of widespread materials for superconducting devices: it is actively used in hot-electron bolometers (HEB) \cite{Lobanov_2015}, superconducting nanowire single-photon detectors (SNSPD) \cite{Vovk_2025}, and photon-number-resolving detectors (PNR) \cite{Marsili_2009}. It is also worth to note niobium-titanium nitride (NbTiN), which has a high critical temperature (up to 15 K) and a high upper magnetic field of $\sim$15 T \cite{Zhang2015, Hazra2018,Sidorova2021}. Due to its properties, NbTiN thin superconducting films have found wide application in various superconducting devices, including SIS-mixers (superconductor-insulator-superconductor) \cite{Wang2013}, superconducting resonators \cite{Endo2013}, and SNSPD \cite{Schuck2013}. Among niobium based intermetallic compounds, NbTi, Nb$_3$Sn, and Nb$_3$Al hold special places. NbTi has $T_c \approx 9,5$ K, and at 4.2 K it features an upper critical field $H_{c2} = 11.5$ T and critical current density 0.4 MA/$\text{cm}^2$. Nb$_3$Sn $T_c \approx 18$ K, at 4.2 K — $H_{c2} = 30$ T and  $J_c \approx 1$ MA/$\text{cm}^2$ \cite{Xu2017A}. Thanks to its characteristics, Nb$_3$Sn is used in areas requiring very strong magnetic fields (above 10-12 T): dipole and quadrupole magnets for accelerators \cite{Dipole2015, Quadrupole2010}, key elements of magnetic systems for thermonuclear reactors (central solenoids and toroidal field coils for ITER-type installations \cite{ITER2015}), as well as MRI magnets \cite{MRI2014}, NMR spectroscopy \cite{NMR2023}. Additionally, developments are underway for using Nb$_3$Sn in superconducting radiofrequency resonators \cite{Resonator2017}. Nb$_3$Al represents a superconducting intermetallic compound with a cubic A15-type structure, characterized by a critical transition temperature to the superconducting state in the range of 17 to 19 K, and an upper critical magnetic field value reaching approximately 25 - 35 T at temperatures approaching absolute zero \cite{Glowacki}. This material is widely used in superconducting magnets and thin-film elements for thermonuclear fusion systems (e.g., ITER) and high-field magnetic systems \cite{Yamada1999}. 

This work presents the results of the synthesis and comprehensive study of a new superconducting compound based on niobium — Nb$_{1-x}$Sc$_x$. Its key characteristics have been determined, including critical temperature ($T_c$), superconducting transition width ($\Delta T_c$), critical current density ($J_c$), upper critical field ($H_{c2}$), diffusivity ($D$), coherence length ($\xi_{GL}$) and a microstructure analysis was carried out.s.  A comparative analysis of the obtained parameters with the properties of classical thin-film superconductors based on niobium (NbN, NbTiN) has been conducted.

%
\section{\label{sec:level1}Fabrication of Nb$_{1-x}$Sc$_x$ films and microbridges}
Silicon wafers of p-type <100> with a diameter of 4 inches were used as substrates, on which a multi-stage cleaning process was carried out in strong acid-alkali solutions. The substrates were then subjected to thermal oxidation to obtain a silicon oxide layer. Nb$_{1-x}$Sc$_x$ films were deposited using the magnetron co-sputtering method from two targets on a Vaccoat DST-3 setup in an argon gas environment. To create a niobium-based compound with a low scandium content, 
two different power sources were used: a direct current (DC) source for sputtering the niobium target and a radio frequency (RF) source for sputtering the scandium target. The choice of different sources is related to the fact that sputtering of materials using an alternating current source is typically much lower than with a constant current source. To determine the optimal percentage ratio of niobium to scandium, a series of thin films was formed on oxide silicon wafers. Within the series of samples, the power density of 
the DC magnetron remained constant at 5.9 $\text{W}/\text{cm}^2$, while the power density of the RF magnetron for different films was 4.9, 7.4, and 9.9 $\text{W}/\text{cm}^2$. The films were deposited at room temperature with a pressure in the process chamber of 6.45 mTorr. The film formation process began with preliminary plasma ignition of Nb and Sc with a shutter closed for several minutes to stabilize the plasma and clean the targets. Figure 1(a) shows the plasma distribution in the vacuum chamber during simultaneous sputtering of Nb and Sc targets. A schematic illustration of Nb and Sc co-sputtering in the vacuum chamber is presented in Figure 1(b).

\begin{figure*}[h]  
    \centering
    \includegraphics[width=1\textwidth]{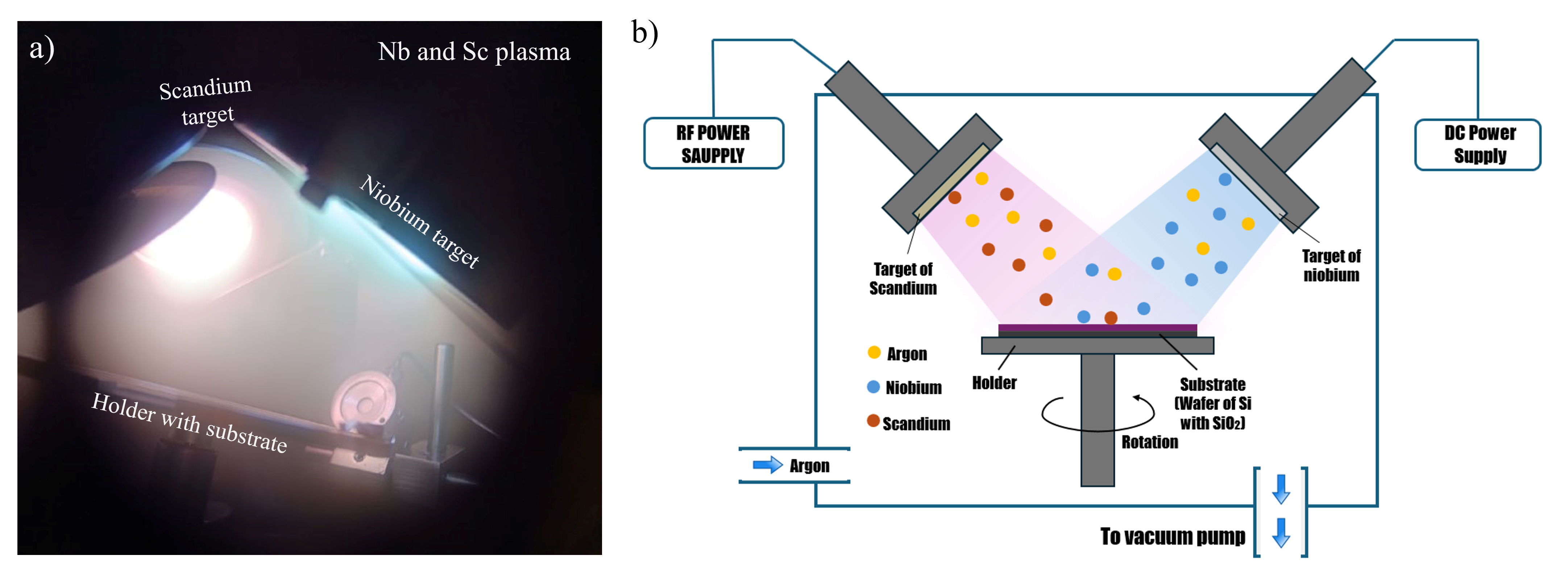}
    \caption{
        (a) Optical photography of plasma during co-sputtering of Nb and Sc targets.
        (b) Schematic illustration of co-sputtering of Nb and Sc.
\centering
    }
    \label{plasma}
\end{figure*} 

To measure the transport and magnetoresistive characteristics of this material,
microbridges were formed. The fabrication process flow 
is shown in Figure 2. The microbridges were formed in 30-nm-thick Nb$_{1-x}$Sc$_x$ layer using a sequence of optical lithography and ion etching operations, resulting in a planar microbridge structure with a length of 50 $\mu$m and width of 2 $\mu$m.

\begin{figure*}[h]
    \centering
    \includegraphics[width=1\textwidth]{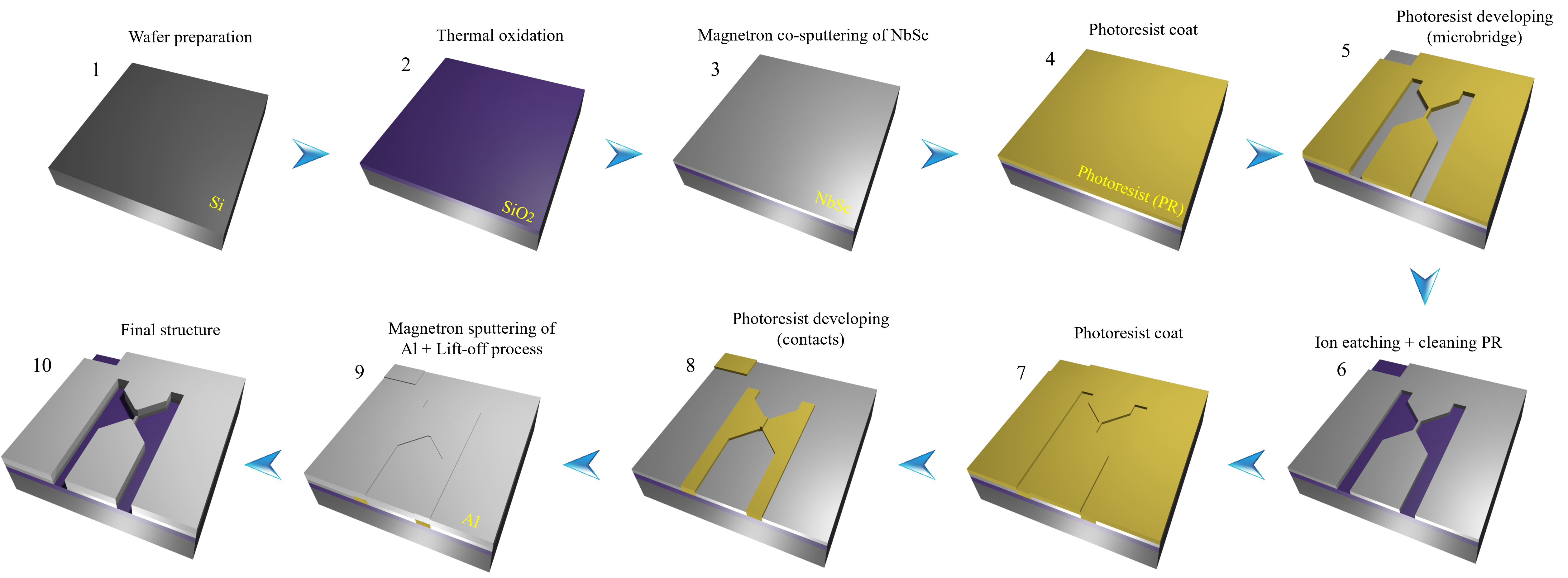}
    \caption{
        The fabrication process flow of microbridge structures based on NbSc.
\centering
    }
    \label{marsh}
\end{figure*}

To ensure a high-quality ohmic contact for measurements of the microbridge structures, aluminum contacts with a thickness of 100 nm were formed in the contact pad areas. Contacts formation was carried out using magnetron deposition technology and lift-off photolithography in organic solvents. Before contacts deposition, surface activation was performed using argon ions to clean and modify the surface of Nb$_{1-x}$Sc$_x$ film
aiming to improve the quality of the Al/Nb$_{1-x}$Sc$_x$ interface. 

\section{\label{sec:level1}Experiment}
Sheet resistance of films at room temperature was measured using the four-probe method. 
The film thicknesses 
were determined using X-ray reflectometry (XRR). Investigation of the elemental composition of Nb$_{0.85}$Sc$_{0.15}$ was performed using Auger electron spectroscopy on the Auger electron spectrometer JAMP-9510F. XRR measurements and structural analysis was carried out using a Malvern Panalytical Empyrean diffractometer with Cu-K$\alpha$ radiation.

The resistance of microbridge structures was measured using the Van der Pauw method. The critical temperature and critical current values of the superconducting transition were measured in a closed-cycle Gifford-McMahon refrigerator cooled to 2.5 K. Thermometry and temperature control were performed automatically using a PID-regulator (LakeShore 325 temperature controller). 
 
The critical current was measured using a two-wire measurement scheme with a precision low-noise current source Keithley 6221A. The temperature dependence of the critical current density was analyzed using current-voltage curves obtained by sweeping the source voltage at a fixed temperature. The critical current density was determined from the critical current value and the cross-section of the microbridge. 

Magnetoresistive measurements were performed in a closed-cycle vertical refrigerator where sample thermalization was achieved using a helium vapor atmosphere. The substrate with the microbridge structure was oriented perpendicular to the applied magnetic field created by a superconducting solenoid in the range of 0 T to 3 T. Measurements were carried out using a four-wire scheme in the temperature interval 1.7 - 7 K at fixed field values.

\section{\label{sec:level1}Results and discussion}
To assess the scandium concentration C(Sc) in Nb$_{1-x}$Sc$_x$ samples at which $T_c$ would be the highest, a series of film samples was investigated using probe spectroscopy and $R(T)$ measurements to determine the critical superconducting transition temperatures. Table I presents the obtained results. The maximum value of $T_c$ was achieved in the sample with approximately 15\% scandium concentration in the film and was 6.35 K. This ratio was selected for further investigation of transport and magnetic properties.
\begin{table}[h]
\centering
\caption{Parameters of the obtained film samples Nb$_{1-x}$Sc$_x$}
\label{tab:Parameters}
\begin{tabular}{c c c c c c}
\hline
Sample & \makecell{Power Density \\ (RF), $\text{W}/\text{cm}^2$} & Thickness, nm & $C(Sc), \% $ & $C(Nb), \% $ & $T_c$, K \\
\hline
1 & 4.93 & 30 & 14.8 & 85.2& 6.35 \\
2 & 7.4 & 32 &16.5 & 32.5 & 5.5 \\
3 & 9.87 & 34 & 18.8 & 81.2 & 4.9 \\
\hline
\end{tabular}
\end{table}

According to the reflectometry results, an atypical ($\omega \approx 1.1-1.4^{\circ}$) form of XRR curve was observed for uniform thickness/density films (Fig. 3) \cite{yasaka2010x}. Among possible explanations for this effect, one can highlight density gradient, film delamination, or lateral thickness non-uniformity. To test the latter hypothesis, XRR curve registration was performed with a smaller aperture (2 mm) to reduce the area of the incident X-ray beam (red curve in Fig. 3). As a result, the oscillations became more monotonic, which agrees with the assumption of material thickness non-uniformity of the film. 

\begin{figure*}[h]
    \centering
    \includegraphics[width=0.5\textwidth]{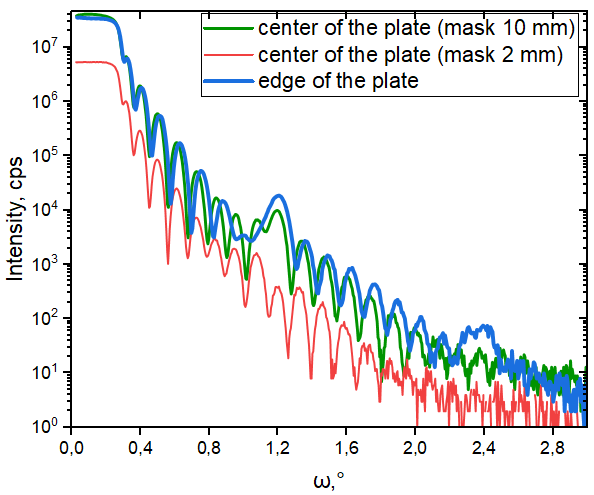}
    \caption{
      XRR curves registered from the edge (blue curve) and center of the wafer using 10 mm (green curve) and 2 mm (red curve) apertures.
\centering
    }
    \label{XRR}
\end{figure*}

The crystal structure of the Nb$_{0.85}$Sc$_{0.15}$ thin film was investigated using grazing incidence X-ray diffraction (GIXRD) at incidence angles $\omega$ in the range from $0.4^{\circ}$ to $1.2^{\circ}$ (Fig. 4(a)). The obtained diffractograms demonstrate the formation of a crystalline phase whose diffraction maximums are close to the cubic Nb phase. However, all observed reflections are systematically shifted toward smaller diffraction angles compared to Nb ICSD card (170906), indicating a significant increase in interplanar spacing and, consequently, lattice parameter expansion. Such behavior can be interpreted as the formation of a substitutional solid solution or a metastable Nb - Sc phase. 

An important feature of the diffraction data is the systematic dependence of the peak position on the incidence angle $\omega$. Thus, the position of the (110) reflection at $\omega = 1.2^{\circ}$ corresponds to $2\theta \approx 37.603^{\circ}$, while at $\omega = 0.4^{\circ}$ the position shifts to $2\theta \approx 37.766^{\circ}$ (inset of Fig. 4(a)). This may be related to a chemical composition non-uniformity and/or a presence of strains in the film. However, depth profiling by electron spectroscopy (ES) did not reveal significant changes in Sc concentration through the film thickness within the experimental error of the method. This observation suggests that the average chemical composition of the film remains practically constant along the growth direction. Therefore, the observed shift of diffraction peaks with incidence angle cannot be explained solely by a change in the chemical composition of the film.

To clarify the role of Sc in the formation of the observed structure, a Sc film was deposited under identical experimental conditions. In its diffraction pattern shown in Fig. 4(b), a broad diffuse halo with weak features near the expected reflections of the hexagonal $\alpha$-Sc phase predominates. The absence of clearly expressed diffraction peaks indicates that Sc deposited under these conditions forms predominantly X-ray amorphous or nanocrystalline film.

\begin{figure*}[h]
    \centering
    \includegraphics[width=1\textwidth]{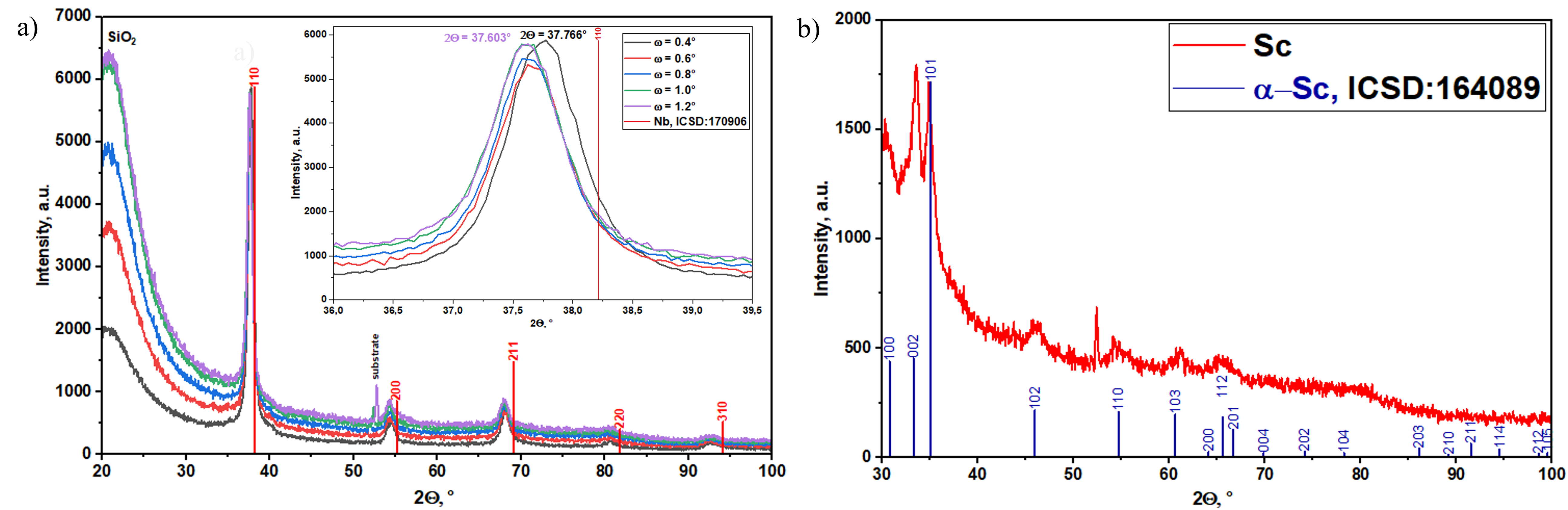}
    \caption{
         (a) Diffractograms of the Nb$_{0.85}$Sc$_{0.15}$ film at $\omega = 0.4^{\circ}/0.6^{\circ}/0.8^{\circ}/1.0^{\circ}/1.2^{\circ}$ with the inset demonstrating the shift in (110) reflection position.
         (b) Diffractogram of the scandium film at $\omega = 1^{\circ}$. Vertical blue lines mark the positions of $\alpha$-Sc phase reflections corresponding to ICSD card 01-077-9986
\centering
    }
    \label{GIXRD}
\end{figure*}

In summary, diffraction data allow concluding that the thin Nb$_{0.85}$Sc$_{0.15}$ film represents a metastable crystalline material with a bcc-like (body-centered cubic) lattice significantly expanded compared to pure Nb. The absence of detected composition gradients indicates that changes in lattice parameter with depth profiling are most likely due to strains in the film.

The distribution of the sheet resistance across the wafer for the Nb$_{0.85}$Sc$_{0.15}$ film is shown in Fig. 5(a). Such a sheet resistance distribution pattern is primarily related to the fact that the material deposition process was carried out from targets with a diameter of 50.8 mm (2 inches) onto a wafer with a diameter of 100 mm (4 inches). The average sheet resistance value is approximately $R_s \approx 31.8 \Omega/\Box$, and the uniformity of distribution across the wafer is $\sim9\%$. The wafer with the fabricated microbridge structures is shown in Fig. 5(b). The microbridge structures were created in the central part of the wafer, where the spread of sheet resistance is smaller. The microbridges were measured using the four-probe method to evaluate their resistance. The geometry of the microbridge structure is characterized by a length-to-width ratio of 50 $\mu$m/2 $\mu$m, which is 25 squares. The average resistance of microbridge structures across the wafer was $\sim$805 $\Omega$.

\begin{figure*}[h]
    \centering
    \includegraphics[width=0.65\textwidth]{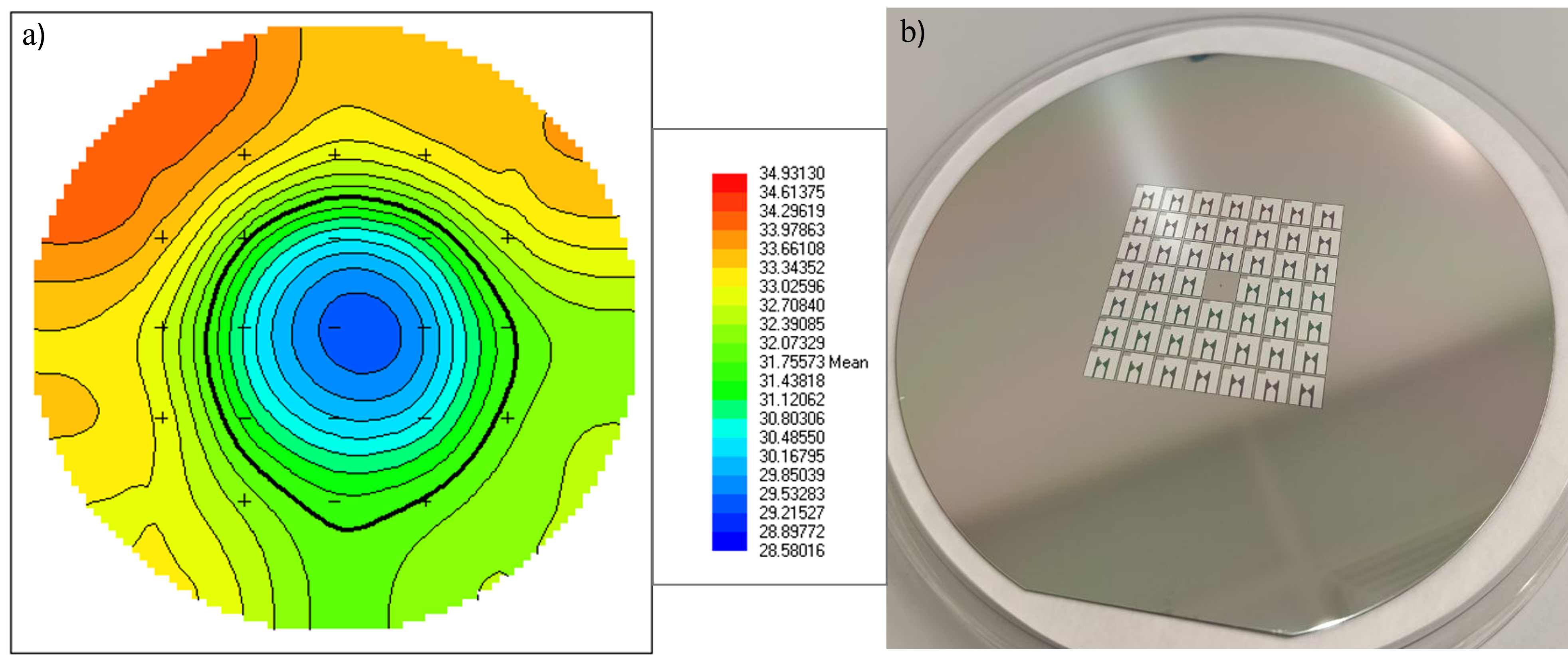}
    \caption{
        (a) Distribution of sheet resistance across the wafer.
        (b) Optical photography of the wafer with microbridge structures. 
\centering
    }
    \label{Rs}
\end{figure*}

For investigation of temperature and transport characteristics, a microbridge sample with a resistance of 854 $\Omega$ at room temperature was selected. Fig. 6(a) shows the temperature dependence of the microbridge resistance 
in the range of 300 K to 2.7 K. The residual-resistance ratio coefficient $RRR$ value, defined as the ratio $R^{300K}/R^{20K}$, equals 1.12. The $T_c$ value was determined as the tempearture at which the maximum of the derivative $dR/dT$ occures in the superconducting transition region and was 6.35 K; the superconducting transition width was estimated as width at half of the maximum of the $dR/dT$ curve and was  $\Delta T_c \approx\text{0.07 K}$.

\begin{figure*}[h]
    \centering
    \includegraphics[width=0.85\textwidth]{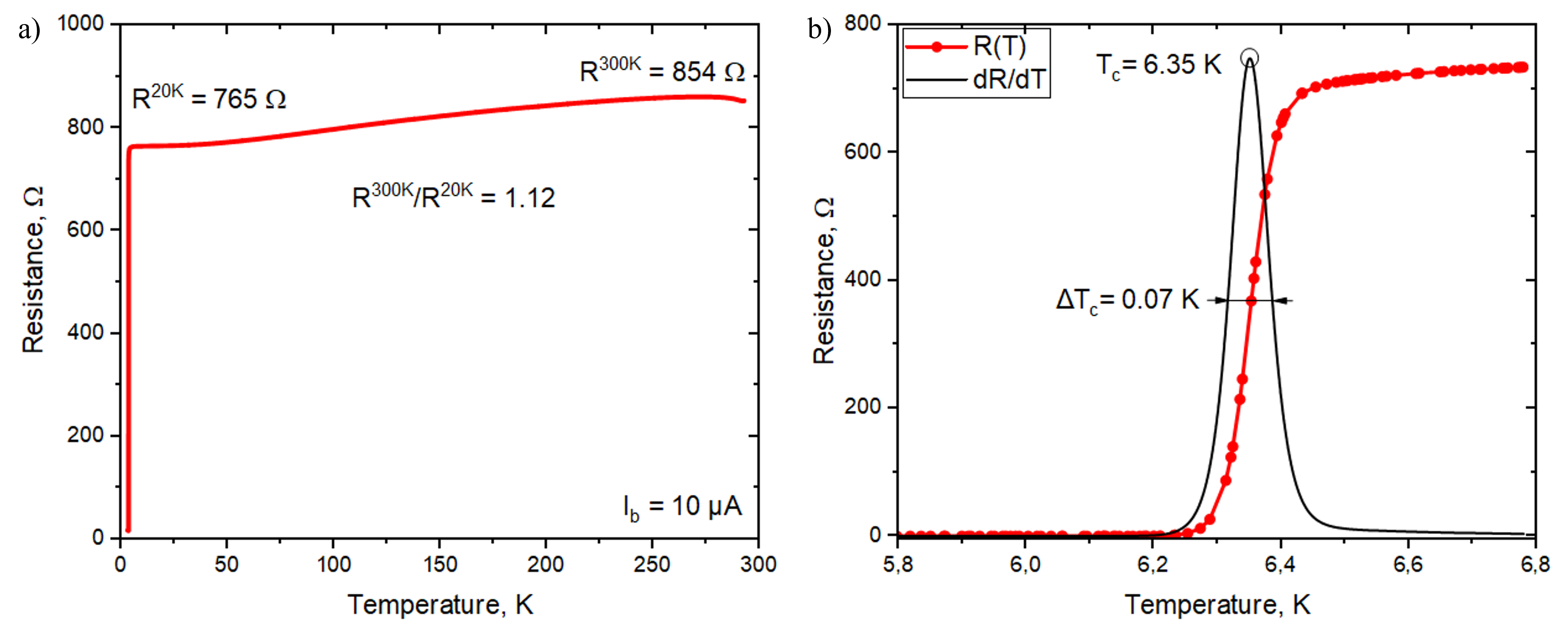}
    \caption{
         (a) Temperature dependence of the sample resistance in the range of 300 K to 2.7 K.
         (b) Temperature dependence of the sample resistance  and its derivative $dR/dT$ in the range of 5.8 K to 6.8 K.
\centering
    }
    \label{R(T)}
\end{figure*}

Fig. 7(a) shows the I-V curve at 2.5 K temperature, from which the critical current density of the microbridge was determined. The slope of the 
superconducting part of the I-V corresponds to 5.3~$\Omega$ resistance of the wires inside the cryocooler. 
The critical current density $J_c$ is be calculated as the ratio of critical current to the microbridge cross-section: $I_{c}(T)/(w \cdot d)$, where $w$ and $d$ are the microbridge width and film thickness respectively. With the obtained critical current value determined as the maximum observable current in the I-V curve, the critical current density $J_{c} \approx 2.5\text{ MA/cm}^{2}$ at 2.5 K.

Fig. 7(b) shows the critical current dependence on temperature. The solid line represents the temperature dependence described by the empirical formula derived from two-fluid model \cite{GL1974}: 
\begin{equation}
I_{c}(T) = I_{c}(0) \left(1-\left(\frac{T}{T_c}\right)^{2}\right)^{3/2} \left(1+\left(\frac{T}{T_c}\right)^{2}\right)^{1/2}
\end{equation}

\begin{figure*}[h]
    \centering
    \includegraphics[width=0.85\textwidth]{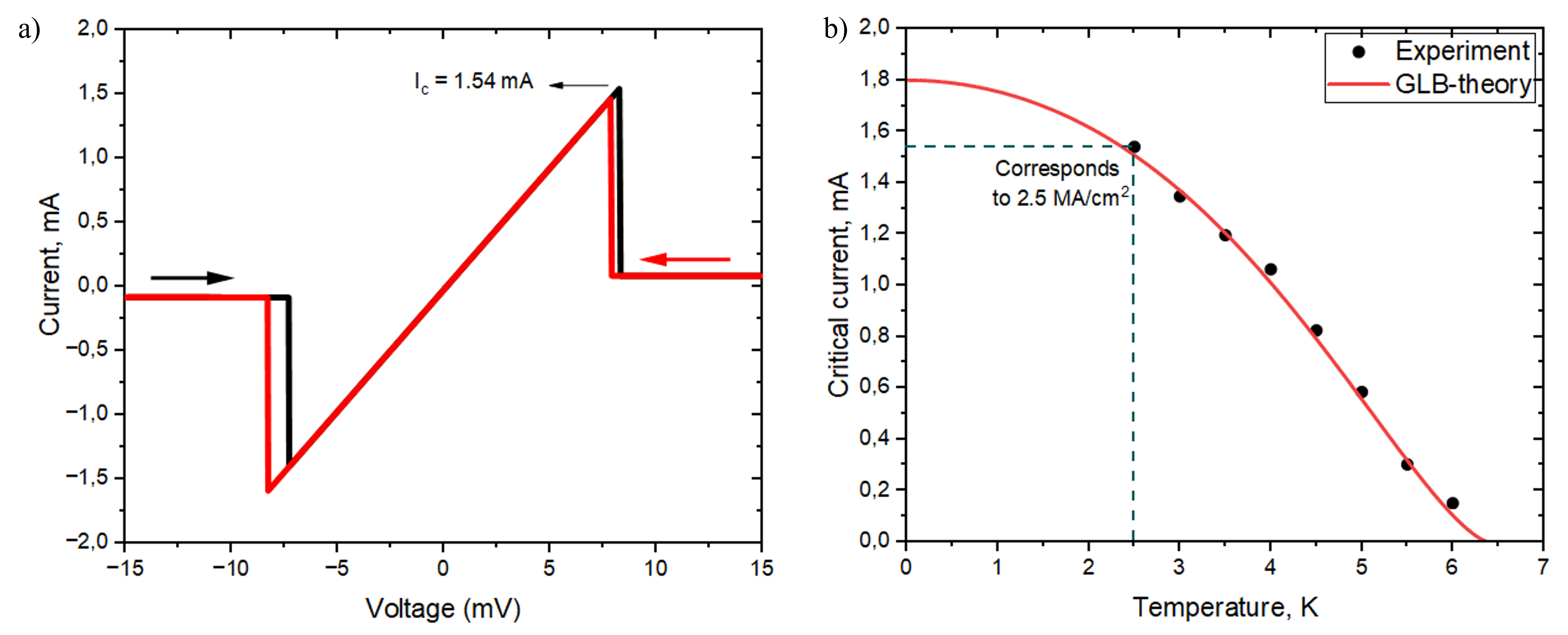}
    \caption{
         (a) Current-voltage characteristic of the Nb$_{0.85}$Sc$_{0.15}$ sample at 2.5 K temperature.
         (b) Temperature dependence of the critical current.
         The solid line shows the temperature dependence of the critical current according to two-fluid model. 
\centering
    }
    \label{Current}
\end{figure*}

To determine the diffusion coefficient $D$ and coherence length $\xi_{GL}$, magnetoresistive measurements of another sample with room temperature resistance of 798 $\Omega$ (about 700 $\Omega$ at 2.5 K) from the same wafer were performed. 
Fig. 8(a) shows $R(T)$ curves measured at various perpendicular magnetic fields. With increasing magnetic field, a decrease in $T_{c}$ and increase in $\Delta T_c$ are observed, which is typical behavior of superconductors under magnetic fields. Based on the obtained dependencies, the second critical field at zero temperature $H_{c2}(0)$ was restored using the following approximation within Ginzburg-Landau theory \cite{Zehetmayer2010}:
\begin{equation}
H_{c2}(T) = H_{c2}(0)\left(1-\left(\frac{T}{T_c}\right)^{2}\right).
\end{equation}

Fig. 8(b) shows the temperature dependence of the magnetic field and the approximating curve with the determined perpendicular field value $H_{c2}(0) = 3.2$ T. The diffusion coefficient was determined from the temperature dependence of the upper critical field at $T = T_c$ using the following expression \cite{Leitner2000}:
\begin{equation}
D = \frac{4\cdot k_b}{\pi \cdot e}\left(\frac{dH_{c2}}{dT}\right)^{-1}.
\end{equation}
As a result, the value $D = 1.1$ $\text{cm}^{2}/s$ was obtained.

\begin{figure*}[h]
    \centering
    \includegraphics[width=0.85\textwidth]{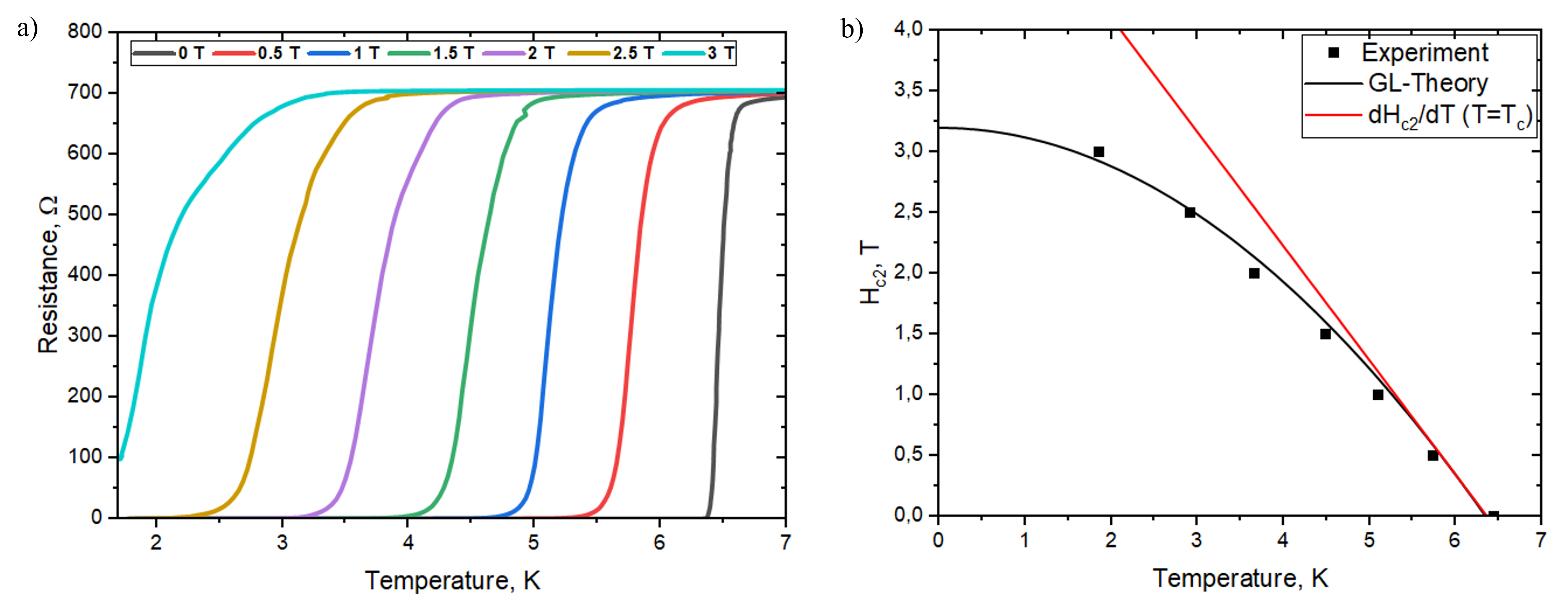}
    \caption{
         (a) Temperature dependences of the resistance of the Nb$_{0.85}$Sc$_{0.15}$ film in the perpendicular magnetic field 
         (b) Experimental dependence of the second critical magnetic field on temperature.
\centering
    }
    \label{magn}
\end{figure*}

The coherence length $\xi_{GL}$ at 0 K was determined using the following formula \cite{Jesudasan2011}:
\begin{equation}
\xi_{GL}(0) = \sqrt{\frac{\Phi_{0}}{2\pi H_{c2}(0)}}.
\end{equation}
where $\Phi_0 = h/2e$ is the quantum of magnetic flux. According to the expression (4), the coherence length value was 10.1 nm. For comparative analysis to determine the application prospects of the new compound, Table II presents the key parameters of three superconducting materials: NbN, NbTiN and studied here Nb$_{0.85}$Sc$_{0.15}$ compound.
\begin{table*}[!h]
\centering
\caption{Parameters of superconducting niobium-based materials}
\label{tab:Parameters of superconducting niobium-based materials}
\begin{tabular}{c c c c c c}
\hline
Material& D, $\text{cm}^{2}/s$& $H_{c2}(0), T$ & $\xi_{GL}, nm$ & $T_{c}, K$ & $ \Delta T_{c}, K$\\ 
\hline
$NbN$ & 0.2 - 1 \cite{NbNDiff} & 10.7 - 22.6 \cite{Sidorova2021,Ilin2014} & 3 - 6.5 \cite{NbNCoh, Hazra2016} & 7 - 15.2 \cite{Porokhov2026,Ilin2014}  & 0.2 - 0.55 \cite{ShibalovM_2022, Hazra2016} \\
$NbTiN$ & 0.7 - 2 \cite{Hazra2018}& 13.9 - 15.2 \cite{Sidorova2021} & 5 - 6 \cite{Hazra2018} & 7 - 15 \cite{Zhang2015, Steinhauer2020} & 0.8 - 1.2 \cite{Steinhauer2020}\\
$Nb_{0.85}Sc_{0.15}$ & 1.1 & 3.2 & 10.1 & 6.35 & 0.07 \\
\hline
\end{tabular}
\end{table*}

The obtained compound exhibits characteristics comparable to the superconducting properties of NbN and NbTiN, indicating its prospects for use in single-photon detectors. The relatively small superconducting transition width compared to other materials may be promising for using Nb$_{0.85}$Sc$_{0.15}$ in transition edge sensors (TES) and hot-electron bolometers (HEB). The low value of the upper critical field allows us to consider the possibility of using the Nb$_{0.85}$Sc$_{0.15}$ in magnetometers.

\section{\label{sec:level1}Conclusion}
As a result of this work, the NbSc compound was obtained and characterized for the first time, synthesized by magnetron co-sputtering from two different sources. Using grazing incidence X-ray diffraction methods, it was established that the observed sample represents a metastable crystalline material with a bcc-like lattice. Measurements of the temperature dependencies of electrical resistance in the temperature range T = 2.7 - 300 K showed that the superconducting state is observed below critical temperature $T_c$ = 6.35 K with the width of the superconducting transition of 70 mK. A material with such a transition width can be used in transition-edge sensors (TES) and hot-electron bolometers (HEB). Additionally, in the work, the critical current density at the temperature of 2.5 K $J_c = 2.5\text{ MA/cm}^{2}$. From the magnetoresistive characteristic measurements we derived the upper critical field $H_{c2}(0) = 3.2$ T, diffusion coefficient $D = 1.1$ $\text{cm}^{2}/s$, and coherence length $\xi_{GL} =  10.1$ nm. The superconducting characteristics of the obtained compound are close to those of NbN and NbTiN, indicating its suitability for single-photon detectors. The low value of the upper critical field, in turn, allows considering this material for magnetometer fabrication.

\begin{acknowledgments}
We express our gratitude to A.V. Goryachev for Auger electron spectroscopy and A.P. Orlov for resistivity measurements. The work was carried out within the framework of the state project of the SRI "Kurchatov Institute". Fabrication and technology characterization were carried out at the large scale facility complex for heterogeneous integration technologies and silicon+carbon nanotechnologies based on Institute of Nanotechnology of Microelectronics of Russian Academy of Sciences.
\end{acknowledgments}

\bibliography{Cites}

\end{document}